# The evolution of Lightning Network's Topology during its first year and the influence over its core values

Stefano Martinazzi[1]

[1]Politecnico di Milano, Department of Management Engineering

It is now a whole year since Lightning Network (LN) has been launched on the Bitcoin's mainnet. LN has been claimed as the solution for several of Bitcoin's weaknesses such as its difficulty to scale in number of transactions per second and its expensiveness for relative small amount exchanged. LN is based upon a network of micro-channels opened and closed by issuing transactions on the Blockchain and capable to interact among themselves thanks to the Multi-hop framework. In this work I analysed the evolution of LN from its topological point of view and tried to understand how it impacts over some of Bitcoin's historical core foundamentals such as the resilience against attacks and failures and the user's anonymity. Another question I have tried to answer is whether the evolution of LN's topology will make its synchronization easier.

## Introduction

Since the its inception, Bitcoin has been known as an instrument incapable of transacting a great amount of transitions per unit of time. Bitcoin has written in its own code that every ten minutes a block can be mined and added to the blockchain[1]. These limitations lead for the major part of Bitcoin's life to have as maximum number 7 transactions per second, to make comparison Visa process routinely 2000 transactions per second, with peaks of several dozens of thousands (1).

Since miners are the one capable to build and add block to the blockchain this low throughput allowed them to impose higher fees in times of great demand. The most emblematic example occurred in 2017 when fees skyrocket from an average less than one dollar per transaction to several dozens. Fees depends on number of transactions waiting to be added in the blockchain, not on the amount of Bitcoins transacted per time, therefore the blockchain can be very cheap compared to traditional means in case of large payments, potentially moving hundreds of millions for few cents, but it is extremely inefficient for routinely payments and for micropayments.

---

[1] Said node must be one megabyte or less.

Several attempts have been tried in order to increase throughputs and lower latencies, for instance in November 2017 a hardfork[2] occurred in the Bitcoin's blockchain with the implementation of Segregated Witness (SegWit) through Bitcoin Improvement Proposal 141[3] (Bip 141) that quadruplicated the number of transactions that is possible to aggregate into a block. Another example occurred in August 2017 with the hard fork that in created Bitcoin Cash, a version of Bitcoin with blocks of 8Mb. Lightning Network is a system of Micropayments channels built on top of Bitcoin's blockchain and therefore referred to as a solution of "layer 2" based on smart contracts. Two nodes can open a channel by issuing a multi signed transaction on the blockchain and thereafter being able to exchange a predefined amount of bitcoins back and forth freely. When the need for payments is fulfilled, another multi-signed transaction can be issued to the blockchain with the final balance. Lightning Network was proposed on February 2015 and in January 2018 the mainnet was launched, after a period of testing on a copy of the original Bitcoin's blockhain called "testned". The Lightning Network is the most recognized solution for scalability. It is based on off chains transactions, meaning that exchanges on the Lightning Network have not to be uploaded to the blockchain at every iteration.

To open a channel a preliminary transaction between to parties is issued on the blockchain, called "channel funding". After that initial, in a sense ordinary, transaction everytime two parties wants to exchange funds on the channel they must issue new "commitment transactions" that are simply the balance of the channel signed by the two parties that has not to be broadcasted to the entire network, with the only exception of the last commitment which is also called "closing transaction" since it closes the channel and set the new balance on the blockchain. If a channel is excessively unbalanced towards one party than said channel is considered "unbalanced", that can constitute a problem for other nodes that are perhaps interested in using that channel to route their transaction through the multi-hop framework. Another important feature of the Lightning Network is the possibility to make payments to nodes not directly connected through a channel via the "multi hop" algorithm that routes payments via nodes eventually reaching the desired one, if possible. Through Hashed Time Lock Contracts (HTLCs) it is possible to send payments to another node without issuing a brand new channel with the condition to find a common path and enough available capacity in each and every channel of it.

During its development one of the most concerning potential issue of Lightning was the possibility of a centralization of the Network. The problem resides in the nature of the multi hop framework. For instance, if node "Bob" wants to send 1 BTC to node "Alice" on the Lightning network, without opening a direct channel, it must search for a node that is connected to both with enough capacity to send its 1 Bitcoin to Alice in exchange of the Bitcoin from Bob, plus a fee. This of course would make nodes with higher capacity more likely to act like a payment hub, de facto centralizing the system. Centralization of Lightning Network would create concerns about users' privacy since hubs would be able to collect information about a huge number of nodes, also hubs could be able to censor transactions or increase their fees (2), (3) and (4).

---

[2] A hardfork is when two parties do not agree on a proposed change in the protocol and they decide thereafter to proceed by themselves on two separate blockchains, one with the change implemented and the other one without.

[3] Bitcoin Improvement Proposal are the standard way for proposing to the community possible enhancements in the Bitcoin protocol.

In the second chapter, I am going to display the dataset used to perform the analysis and some preliminary information about how I decided to construct the network. In the third part of this article, I will present some topological information and their evolution through a period of 12 months. The fourth section is dedicated to analyse LN in the light of three different but extremely importance aspects, its robustness to failures and malicious attacks, its topological capability to help in the preservation of anonymity and finally if its topology is effective in promoting an environment where synchronization can be achieved.

## Dataset and the Network

Retrieving data for analysing the Lightning Network can be quite challenging. Transactions used to open and close channels have been developed to be virtually indistinguishable from other smart contracts issued on the blockchain. The reason is that Miners, which are rewarded both with newly minted Bitcoin and with transaction fees, have the interest to keep the flow of transactions on the blockchain and they could prevent channels from opening by simply never including those transactions, if they were able to recognize them. In order to be able to obtain data from the lightning network we should open a channel with some highly connected node. Another way is to retrieve the information directly from pages that make their knowledge about the network available. We opted for this second option, in particular we used data from hashxp.org and engine for search and analysis in the lightning network.

It is important to note that it is impossible to know if the data in our possession is the totality of the network or if there are some other nodes not directly or thorough Multi-Hop connected to the ones in 1ml.com possession. Using R's packages Rvest and Jsonlite we retrieve data spawning from 12 of January 2018, launch of LN on the mainnet, to 12 of January 2019. Exchange rates between Bitcoin and USD and average transaction fees on the blockchain are taken from bitinfocharts.com.

The final dataset has been composed by twelve different snapshots corresponding to the twelve of each month since February 2018 Fig. [1.a] until January 2019 Fig. [1.b]. The Biggest connected components for each month accounted for the entirely of the network, with only few disconnected components composed just by single couplets[4].

---

[4] We must consider that there are components that are not known by us, but it is logical to assume that their number and sizes are absolutely dwarfed by our component, which was built on the base of information collected from one of the most important LN's operator.

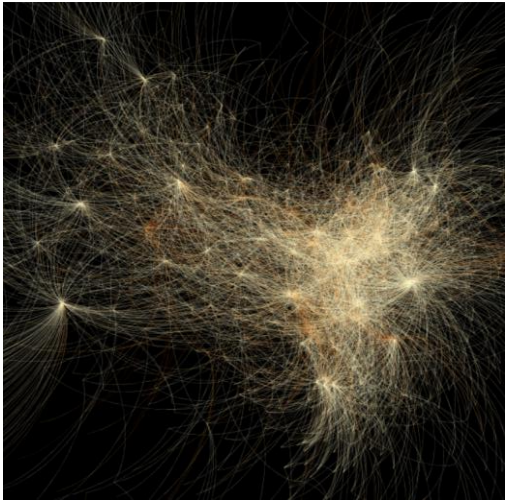 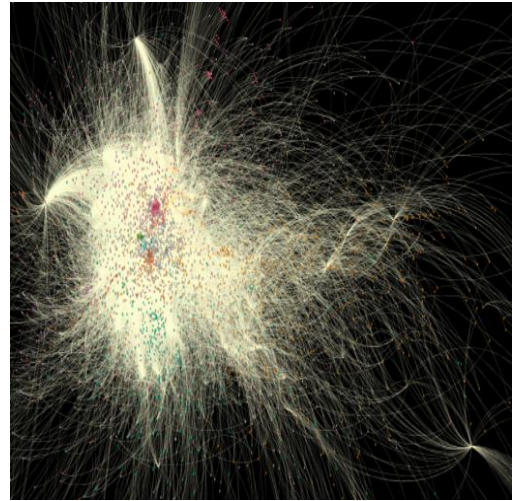

*Figure 1.a LN 02/12/2018*    *Figure 1.b LN 01/12/2019*

In June 2018 Tony Arcery showed on twitter how the probability of successfully routing a transaction through the Lightning Network is 100% or so just for amounts below 0.001 BTC, after that threshold said chances decrease dramatically reaching zero at less than 0.1 BTC.

This is of course linked to the median capacity actually installed among LN's channels, this suspicion has its origin on the fact that Mr. Arcery performed the experiment in February and again in June and the two results followed quite truthfully the trend in the distribution of the number of nodes per capacity installed (5). For this reason I decided to perform most of the analysis using a weighted undirected network, undirected because there is no way to know one by one the distribution of funds on channels, therefore we assume for simplicity they are always balanced.

As weights, we used the reciprocal of the installed capacity expressed in USD, in order to mitigate for the extreme volatility of Bitcoin's price during all 2018.

## Topological evolution of Lightning Network

In this chapter, I am going to provide with some preliminary descriptions of how LN has evolved during this year. The number of node connected at the time of each moment grew from 518 to 3613 and the number of channels from 1910 to 23860, therefore the Network evolved from an already remarkable sparsity even more so. Since its establishment LN has assumed the scale-free configuration, with the exponential ranging around 2, which is confirmed by the Kolmogorov-Smirnov test run for every take, as in table 1.

**Table 1: Kolmogorov-Smirnov test results**

|        | Alpha  | KS_test | p_val  |
|--------|--------|---------|--------|
| feb-18 | 1.9415 | 0.0472  | 0.7693 |
| mar-18 | 2.0474 | 0.0203  | 0.9996 |
| apr-18 | 2.1397 | 0.0344  | 0.8098 |
| mag-18 | 2.1048 | 0.0273  | 0.8310 |
| giu-18 | 2.1139 | 0.0239  | 0.9391 |
| lug-18 | 2.0443 | 0.0311  | 0.5763 |
| **ago-18** | 2.0705 | 0.0352 | 0.5361 |
| **set-18** | 2.0917 | 0.0374 | 0.6040 |
| **ott-18** | 2.1684 | 0.0313 | 0.8854 |
| **nov-18** | 2.1341 | 0.0394 | 0.5890 |
| **dic-18** | 2.0764 | 0.0358 | 0.5476 |
| **gen-19** | 2.0694 | 0.0398 | 0.2628 |

The median degree, a measure quite important if we consider both the necessity to routing transactions but also the vocation of Bitcoin to not be decentralised and for this reason more meaningful than the average degree, increased doubling from a median of 2 to 4 connections per node.

Since I decided to consider LN as weighted, it essential also to look for the median strength of nodes and its evolution. The strength of a node is the sum of all its edges weights, taking into consideration the fact that weights are normally considered as costs to pass through an edge, and our are the opposite, a lower strength means an higher capability of that node to accept flows of transactions through its channels.

The Network present a visibly diassortativity, nodes with lower degree are easier to open channels with others with higher connectivity, this could be a signal of the formation of hubs in the network. This sort of behaviour is reasonable, since opening channels is costly the multi-hop routing system, thorough an highly connected node, can be a way cheaper solution in order to reach a large number of other vertexes.

If, instead of degree assortativity, I compute the assortativity for the weighted graph, as proposed by Leung and Chau (6), than we can see how the diassortative tendency of the Network disappears. This behaviour can be explained if we compute the correlation between nodes' degree and their average capacity per channel. Surprisingly we can notice how being highly connected is not strongly correlated with the average capacity.

This phenomenon might be explained by the fact that "poor" nodes prefer to connect directly to hubs in order to save transaction fees to open and close channels and routing fees, while more "wealthy" nodes have not this concern and connect directly to every node they are interested skipping hubs.

This could pose a threat and a drawback from Bitcoin's core values, though. An highly connected node can be used to harvest a great amount of information coming from the flow it is able to intercept, it can also censor transactions from or to a certain node, even if the sender change the routing plan, since there is an high probability that the malicious hub is connected to another of the passing nodes.

Even if the hub is legit, its presence could constitute a liability to the LN since it poses as a preferential target to attacks deemed to destabilize the network, and for a considerable amount of time since if not indefinitely depending on the surface of aggression exploited[5], this matter is going to be discussed deeper in the next section.

---

[5] In March, BitPico sent down the 20% of nodes with a DDoS attack, another strategy would be to steal the private keys and close pivotal channels. In that case, and in the best scenario, it would be necessary to wait at least for ten minutes before starting recreating the initial set of connections.

Finally I studied how the efficiency of the network, i.e. its ability to transfer information through its nodes with the fewer number of hops necessary, in other words the ratio between the efficiency of a network where every node is linked to another and the sum of the reciprocals of all shortest paths among nodes.

Efficiency can be computed only taking into consideration nodes and their links or also the weighted Network, and we see how the unweighted version shows no improvement whatsoever while the weighted one improved sensibly month after month[6]. This is very likely due to the fact the capacity installed increased continuously during this period.

Table 2: A collection of topological measures about LN

|  | Nodes | Channels | Median degree | Median Strenght | Total cap in BTC | Total cap in USD |
|---|---|---|---|---|---|---|
| **feb-18** | 518 | 1910 | 2 | 1.08 | 6.24 | 54,072 |
| **mar-18** | 733 | 2060 | 2 | 1.72 | 4.70 | 44,401 |
| **apr-18** | 1359 | 6029 | 3 | 3.93 | 14.96 | 110,003 |
| **mag-18** | 1721 | 8172 | 3 | 2.55 | 20.80 | 175,503 |
| **giu-18** | 1808 | 7876 | 3 | 3.19 | 23.24 | 157,455 |
| **lug-18** | 2039 | 8996 | 3 | 3.06 | 62.07 | 388,082 |
| **ago-18** | 2130 | 11137 | 3 | 3.17 | 95.19 | 601,241 |
| **set-18** | 2337 | 12312 | 3 | 1.61 | 115.47 | 725,934 |
| **ott-18** | 2466 | 12429 | 3 | 1.59 | 113.58 | 713,085 |
| **nov-18** | 2626 | 12958 | 3 | 1.03 | 114.16 | 733,584 |
| **dic-18** | 2878 | 17086 | 3 | 1.34 | 469.63 | 1,635,724 |
| **gen-19** | 3613 | 23860 | 4 | 0.86 | 581.42 | 2,121,019 |
|  | Assortivity (weig.) | Assortivity (unweig.) | Efficiency | Transitivity | Min. Spaw. Tree | Density |
| **feb-18** | -0.051 | -0.370 | 8.94 | 0.12 | 0.269 | 1.4% |
| **mar-18** | -0.110 | -0.369 | 4.76 | 0.05 | 0.351 | 0.8% |
| **apr-18** | -0.143 | -0.267 | 5.99 | 0.09 | 0.222 | 0.7% |
| **mag-18** | -0.129 | -0.293 | 7.75 | 0.09 | 0.209 | 0.6% |
| **giu-18** | -0.140 | -0.284 | 6.38 | 0.07 | 0.227 | 0.5% |
| **lug-18** | -0.107 | -0.263 | 10.35 | 0.07 | 0.225 | 0.4% |
| **ago-18** | -0.014 | -0.249 | 17.37 | 0.09 | 0.190 | 0.5% |
| **set-18** | -0.025 | -0.258 | 24.93 | 0.09 | 0.188 | 0.5% |
| **ott-18** | -0.035 | -0.252 | 23.21 | 0.09 | 0.197 | 0.4% |
| **nov-18** | -0.080 | -0.269 | 22.79 | 0.08 | 0.201 | 0.4% |
| **dic-18** | -0.022 | -0.240 | 42.65 | 0.10 | 0.167 | 0.4% |
| **gen-19** | -0.006 | -0.220 | 42.18 | 0.10 | 0.150 | 0.4% |

## LN Robustness, Anonymity and Synchronizability

Because of the presence of channels with different capacities stored on them, LN can be seen in the same way as an infrastructure for transporting energy. Taking into consideration only the unweighted topology of the Network would possibly lead to erroneous conclusions about its robustness and capability to withstand an aggression. This has been showed in a recent paper by Bellingeri and Cassi, the framework of which is going to be my own (7).

---

[6] Be aware that we should not compare weighted version of a measure against the unweighted one but only appreciate their own evolution in time.

In their study they used the size of the largest connected component to assess the unweighted entity of the damage suffered after a failure and the Delta Weighted Efficiency Loss to assess the weighted size. They performed these measures taking into consideration several attacking strategies and the random event, which may occur in case of a non mischievous faultiness.

For my research I am going to implement as first attacking strategies removing high degree node first, the second using the between centrality and their weighted counterparties. I could consider the number of nodes attacked as a function of the LN's size or letting it constant considering the resources of an attacker do not depend on LN's size, I opted for the latter solution and performed the attacks.

All typologies of attacks happen both on an unweighted graph and with weights, removing 1, 2, 5, 10, 25 and 50 nodes each time. Random failures' number is surely linked with the size but, for sake of comparability with actual attacks, I opted for the same constants.

The most effective strategies were based on between centralities, with the weighted being slightly better. We can see in table 3 how, after a sudden raise between February and March, the damage suffered by the network constantly decrease. Overall, the LN seems to be very resistant to random faultiness with drops in efficiency above 10% just in few occasions.

**Table 3: Efficiency drops in weighted Bet. Centrality attacks**

| Nodes removed | feb-18 | mar-18 | apr-18 | mag-18 | giu-18 | lug-18 |
|---|---|---|---|---|---|---|
| 1 | -8% | -38% | -19% | -29% | -17% | -14% |
| 2 | -12% | -41% | -41% | -33% | -29% | -26% |
| 5 | -23% | -51% | -56% | -45% | -48% | -34% |
| 10 | -34% | -67% | -74% | -59% | -54% | -47% |
| 25 | -50% | -90% | -87% | -82% | -74% | -66% |
| 50 | -66% | -96% | -96% | -91% | -85% | -81% |
|  | **ago-18** | **set-18** | **ott-18** | **nov-18** | **dic-18** | **gen-19** |
| 1 | -10% | -31% | -22% | -16% | -13% | -6% |
| 2 | -24% | -36% | -39% | -19% | -29% | -9% |
| 5 | -35% | -45% | -53% | -44% | -42% | -19% |
| 10 | -49% | -57% | -62% | -54% | -49% | -29% |
| 25 | -58% | -69% | -75% | -69% | -66% | -52% |
| 50 | -79% | -82% | -85% | -80% | -79% | -68% |

**Table4: Efficiency drops in weighted random failures**

| Nodes removed | feb-18 | mar-18 | apr-18 | mag-18 | giu-18 | lug-18 |
|---|---|---|---|---|---|---|
| 1 | -5% | 0% | 0% | -6% | -9% | -4% |
| 2 | -5% | -1% | 0% | -6% | -9% | -4% |
| 5 | -5% | -2% | 0% | -6% | -9% | -4% |
| 10 | -5% | 0% | -3% | -6% | -9% | -5% |
| 25 | -5% | -7% | -2% | -7% | -11% | -6% |
| 50 | -6% | -1% | -10% | -7% | -11% | -12% |
|  | **ago-18** | **set-18** | **ott-18** | **nov-18** | **dic-18** | **gen-19** |
| 1 | -3% | -5% | -5% | -5% | -4% | -3% |
| 2 | -4% | -5% | -5% | -5% | -5% | -3% |
| 5 | -3% | -5% | -5% | -5% | -4% | -3% |
| 10 | -7% | -6% | -6% | -5% | -4% | -3% |
| 25 | -6% | -8% | -7% | -6% | -5% | -3% |
| 50 | -9% | -5% | -6% | -18% | -5% | -3% |

As I mantioned earlier, following the work of Bellingeri and Cassi, I used also the percentage of the largest connected component that converged into the new one after the attacks/failures. Also in this case the most efficient way to maliciously meddling with LN is through attacks based on weighted between centrality.

This second measure confirms what have been previously noted, the evolution of LN is making it somewhat stronger against highly directed attacks and it prooves to be very robust in case of random non hostile failures. Is interesting to notice how a single malicious remotion of a node do not compromise the unweighted topology of the network but provokes a slightly reduction in its efficiency, this should be interpret as a clear signal of few redundancies with sufficient capacity to provide one with another.

This has some major implication not only for attacks and failures but also for normal routine such as the deliberate closure of a channel by its owner, which may result in having the same effect of a non-wanted event. This will be adressed in this chapter taling about synchronization and coordination among nodes.

**Table 5: Percentage of the original LCC remained after weighted Bet. Centrality attacks**

| Nodes removed | feb-18 | mar-18 | apr-18 | mag-18 | giu-18 | lug-18 |
|---|---|---|---|---|---|---|
| 1 | 100% | 86% | 92% | 96% | 100% | 99% |
| 2 | 98% | 85% | 82% | 96% | 96% | 96% |
| 5 | 98% | 84% | 78% | 94% | 91% | 95% |
| 10 | 94% | 73% | 70% | 87% | 90% | 90% |
| 25 | 89% | 58% | 58% | 72% | 82% | 85% |
| 50 | 82% | 32% | 35% | 62% | 70% | 71% |
| | ago-18 | set-18 | ott-18 | nov-18 | dic-18 | gen-19 |
| 1 | 99% | 100% | 100% | 100% | 100% | 100% |
| 2 | 95% | 99% | 99% | 100% | 98% | 100% |
| 5 | 94% | 95% | 95% | 96% | 96% | 96% |
| 10 | 89% | 93% | 91% | 94% | 91% | 93% |
| 25 | 85% | 86% | 84% | 85% | 85% | 87% |
| 50 | 74% | 75% | 73% | 73% | 76% | 78% |

**Table 6: Percentage of the original LCC remained random failures**

| Nodes removed | feb-18 | mar-18 | apr-18 | mag-18 | giu-18 | lug-18 |
|---|---|---|---|---|---|---|
| 1 | 100% | 100% | 100% | 100% | 100% | 100% |
| 2 | 100% | 100% | 100% | 100% | 100% | 100% |
| 5 | 100% | 99% | 99% | 99% | 100% | 100% |
| 10 | 100% | 98% | 98% | 99% | 99% | 98% |
| 25 | 99% | 94% | 96% | 98% | 98% | 98% |
| 50 | 98% | 90% | 92% | 96% | 97% | 96% |
|  | ago-18 | set-18 | ott-18 | nov-18 | dic-18 | gen-19 |
| 1 | 100% | 100% | 100% | 100% | 100% | 100% |
| 2 | 100% | 100% | 100% | 100% | 100% | 100% |
| 5 | 100% | 100% | 100% | 100% | 100% | 100% |
| 10 | 99% | 99% | 99% | 100% | 100% | 100% |
| 25 | 99% | 98% | 99% | 99% | 99% | 99% |
| 50 | 97% | 98% | 96% | 96% | 98% | 98% |

Bitcoin has always been linked with anonymity, before becoming a highly speculative asset it achieved fame mostly due to its use in the dark market. Some attempts to de-anonymize components of a network can rely upon its topological characteristics and weakness.

In preserving the privacy on nodes a role is played also by the Network's configuration itself, for example edge k-anonymity is a technique that sever connections and creates new one in order to arrange groups of nodes indistinguishable with respect to at least k-1 other peers. Singh and Zhan proposed a metric to assess the level of anonymity provided by the graph itself called "topological anonymity" (ta) which is a composite indicator derived from degrees and clustering coefficients (8).

$$ta = \frac{\sum_{i=1}^{\max(\deg(G))}(|Di| * CC_{difi}) - \sum_{j=1}^{\epsilon-1}|Dj|}{n}$$

with |Di| the number of nodes with degree D and CC the Boolean cluster coefficient, which is 1 if var(CC(D))>0 and zero otherwise. In their paper, they assessed ta of different networks and with different levels of security required, expressed with the parameter $\epsilon$ from 2 to 4.

**Table 7: Lightning Network's ta evolution**

| 02/2018 | 03/2018 | 04/2018 | 05/2018 | 06/2018 | 07/2018 |
|---|---|---|---|---|---|
| 0.34 | 0.24 | 0.43 | 0.44 | 0.41 | 0.43 |
| 08/2018 | 09/2018 | 10/2018 | 11/2018 | 12/2018 | 01/2019 |
| 0.46 | 0.46 | 0.45 | 0.44 | 0.47 | 0.49 |

Taking again into consideration that privacy is perhaps the raison d'etre of Bitcoin, I decided to set the level on 4, the maximum. From my analysis, we can see how LN is actually improving its overall topological resistance to anonymity breaches of nodes and edges.

Bitcoin is a decentralized system where consensus and coordination are forced using the "Proof of Work" algorithm, on Lightning Network there is not such system and the behaviour of its components is left to their own ability, if any, to coordinate among themselves.

Coordination in LN is particularly important for those nodes that we have seen playing critical roles in the network. Eventually those nodes will have to close channels in order to upload the new balance on the blockchain, for any kind of reason one of which might be the channel is too unbalanced. If some channels were closed, the best scenario would consist of a network fully capable to offsets for all the time necessary until the connection is established again.

Synchronizability is a critical feature in Networks for them to be able to coordinate themselves and Pecora and Carroll proved that one of the key factor lies within the Laplacian Matrix's eigenvalues, in particular in the ration between the second and the smallest non zero ones, known as the eigenratio.

In particular the lower is the value of the eigenratio the more the network is synchronizable and viceversa (9). My results shows that after a sudden increase in the eigenratio between mid-February and mid-march, similar to what we seen with the robustness analysis, its value continued to growth but at a slower pace, indicating a degradation in LN's synchronizability which can lead to disservices during some pivotal channel's closing-reopening.

In fact, we could represent the entire LN as an ensemble of multi-state oscillators, the channels, with three different possible states: "Open & Balanced"; "Open & Unbalanced" and "Closed". A channel could arrive in the closed and open & balanced states from both the other two, while the unbalanced state can be reached only from the balanced one.

Since it is impossible to know the distribution of funds in all channels without reaching them directly through the multi-hop, a more syncronizable topology would help in reducing the difficulty to collect this kind of information and moving from a state to another accordingly both with the node's own necessities as well with the Network's ones.

**Table 8: Lightning Network's Eigenratio evolution**

| feb-18 | mar-18 | apr-18 | mag-18 | giu-18 | lug-18 |
|---|---|---|---|---|---|
| 7.82 | 24.00 | 24.18 | 25.71 | 24.81 | 25.83 |
| **ago-18** | **set-18** | **ott-18** | **nov-18** | **dic-18** | **gen-19** |
| 25.45 | 25.54 | 25.57 | 25.78 | 29.92 | 29.94 |

## Discussion and Conclusions

In this paper I presented you the topological analysis of Bitcoin's Lightning Network performed through all its first year of existence on the mainnet.

During this period this period the number of nodes increased by almost 7 times and the number of channels simultaneously available by more than 12 times. The value loaded on channels is still negligible if compared with Bitcoin's $60 billion market cap as of the time I am writing, but is growing rapidly both in total value as well as in average per channel.

I showed also how the fear that LN would have brought with itself centralization was not without basis, since it now clearly presents a structure of high degree hubs to whom low degree nodes prefer to attach in order to be able to reach more other peers without having to establish direct connection with them.

I show that this process do not affect richer nodes that prefer to bypass hubs connecting directly with whoever they want, we can imagine this situation as a contraposition of malls for average people against boutiques for the wealthier.

I have also investigated LN from three point of view, for the best of my knowledge, quite critical for it success, which are its robustness, its ability to preserve privacy and to enhance coordination between its nodes.

We have seen how the network autonomously is improving its resistance both to random non malicious failures as well as against attacks and how the evolution of its topology is improving, from a topological point of view at least, the preservation of the users' privacy.

Finally, the part that seen LN growth somehow wanting was its ability to create the condition for reaching coordination, which is constantly dropping. This dynamic could pose a serious threat for a wider adoption of this solution, since the correct functioning of the entire network relies on the shoulders of few nodes and their channels.